\pdfoutput=1
\documentclass[11pt]{article}

\usepackage[utf8]{inputenc}
\usepackage[T1]{fontenc}
\usepackage{textcomp}                 
\usepackage[margin=1in]{geometry}
\usepackage{graphicx}
\usepackage{array}                    
\usepackage{booktabs}                 
\usepackage{tabularx}                 
\usepackage{amsmath,amssymb}
\usepackage{authblk}                  
\usepackage[colorlinks,urlcolor=blue,linkcolor=blue,citecolor=blue]{hyperref}
\expandafter\def\expandafter\UrlBreaks\expandafter{\UrlBreaks\do\/\do\*\do\-\do\~\do\'\do\"\do\-}

\title{Power-Flexible AI Data Centers: A New Paradigm for Grid-Responsive Compute\thanks{This
  work has been submitted to the IEEE for possible publication. Copyright may be transferred
  without notice, after which this version may no longer be accessible.}}

\author[1]{Chris Williams}
\author[1]{Philip Colangelo}
\author[1]{Ayse Coskun}
\author[1]{Ethan Levine}
\author[1]{Andy Neale}
\author[1]{Ciaran Roberts}
\author[1]{Shayan~Sengupta}
\author[1]{Nikhil Shirolkar}
\author[1]{Varun Sivaram}
\author[1]{Sarah Soares}
\author[1]{Ethan Tiao}
\author[1]{Scott Underwood}
\author[1]{Daniel Wilson}
\author[2]{Frank Sharp}
\author[3]{Luke Wainwright}
\author[4]{Harry Petty}
\author[4]{Scott Wallace}
\author[5]{Brandon Records}

\affil[1]{Emerald AI}
\affil[2]{Electric Power Research Institute (EPRI)}
\affil[3]{National Grid}
\affil[4]{NVIDIA}
\affil[5]{Oracle}

\date{}

\begin{document}

\maketitle

\begin{abstract}
The rapid expansion of artificial intelligence (AI) infrastructure is driving unprecedented growth in electricity demand from data centers. Traditional power-system planning treats large computing facilities as inflexible peak loads, leading to costly infrastructure upgrades and long delays in grid interconnection. Recent work has shown that AI clusters can reduce electricity consumption during peak demand through software-based workload orchestration. This article explores how modern GPU-based AI data centers can operate as grid-interactive assets that respond dynamically to power system conditions. We describe an architecture integrating grid signals, workload scheduling, and power telemetry for fine-grained cluster power control. Experimental results from a real-world deployment on a 130 kW GPU cluster demonstrate multiple forms of flexibility, including rapid load reduction, sustained curtailment, and carbon-aware operation while preserving service levels for priority jobs. We further demonstrate performance-aware load shifting across geographically distributed clusters, enabling workloads to migrate toward regions with lower grid stress. Together, these capabilities transform AI infrastructure from static electricity consumers into flexible resources that support grid reliability, accelerate interconnection, and improve computing sustainability.
\end{abstract}

\section{Introduction}

AI is rapidly transforming the global digital economy, but its infrastructure footprint is growing just as quickly. Large-scale AI training clusters and inference services now require massive computing resources, and the electricity demand associated with these systems is increasing dramatically. Modern AI factories, which are large GPU-based computing facilities dedicated to training and operating advanced models, can consume tens to hundreds of megawatts of power, with future facilities expected to reach gigawatt scale \cite{iea}.

This rapid growth presents a challenge for power systems worldwide. Grid infrastructure often evolves on timescales measured in years or decades, while AI infrastructure commonly expands on software and semiconductor development cycles measured in months. In many regions, the resulting mismatch between computing demand and grid capacity has created significant bottlenecks in the process of connecting new data centers to the power grid. Transmission upgrades, new substations, and additional generation resources often require lengthy planning, permitting, and construction processes \cite{iea}.

A central reason for these delays is how power systems traditionally model large industrial loads.
Grid planners typically assume large data centers as firm loads that must receive full power during all operating conditions. This assumption leads to costly infrastructure upgrades and long interconnection delays \cite{luo}.


This article explores methods for making AI data centers power-flexible.
Recent advances in software-defined infrastructure suggest that traditional firm load assumptions may no longer be necessary. 
Large GPU clusters typically run a mix of AI workloads, including long-running training jobs, fine-tuning tasks, batch inference pipelines, and interactive inference services. While some tasks are latency sensitive, many others are throughput-oriented and can tolerate short-term variations in execution speed or scheduling \cite{chen}.
These characteristics create opportunities for modulation of data center power consumption without interrupting critical services. Software systems that orchestrate AI workloads can adjust job scheduling, apply GPU power caps, or shift workloads across time and location. These capabilities allow computing infrastructure to dynamically adapt its power consumption in response to grid conditions.

Recent field demonstrations have begun to illustrate the potential of this approach. In a recent trial, an AI cluster was able to reduce its electricity consumption by 25\% during peak demand periods while maintaining service levels for jobs \cite{colangelo}. Such results suggest that AI infrastructure could function not merely as a passive consumer of electricity, but as an actively controllable load capable of supporting grid reliability.

In many power systems, to fully realize this potential, data centers must evolve beyond simple peak-demand curtailment. Grid-interactive computing systems must be capable of responding to real-time dispatch signals based on established utility protocols, sustaining power reductions for extended periods, coordinating across multiple facilities, and aligning energy consumption with renewable availability and carbon intensity signals.
This article introduces the data center software architecture and capabilities required to enable such systems. Specifically, we describe a software-defined framework for \textbf{power-flexible AI infrastructure}, where computing facilities dynamically adjust their electricity consumption based on grid conditions while preserving application service levels.

The contributions of this article are as follows:
\begin{enumerate}
    \item \textbf{Architecture}: We present an integrated architecture for power-flexible AI data centers combining grid interfaces, workload orchestration, and real-time power modeling.

    \item \textbf{Experimental demonstration}: We analyze results from experimental deployments showing that GPU clusters can deliver multiple forms of grid services including rapid load reduction and sustained curtailment.

    \item \textbf{Carbon-aware and geo-distributed computing}: We introduce mechanisms that align AI workload scheduling with carbon intensity and grid conditions across multiple data centers.
\end{enumerate}

Experimental demonstrations in the US and UK have validated the capabilities of software-defined workload orchestration for delivering high-granularity power flexibility and geo-spatial load shifting in response to real-time grid signals:
\begin{itemize}
    \item \textbf{UK Grid Compliance}. Achieved 100\% compliance across 200+ distinct National Grid power events on a 130 kW AI cluster of NVIDIA Blackwell Ultra GPUs.
    \item \textbf{US Inference Geo-Shifting}. Successfully migrated 10\% of live inference traffic from an Oracle GPU cluster consisting of NVIDIA H100s in Virginia to Illinois during a Dominion Energy utility winter peak power event.
\end{itemize}

Following these innovations, we discuss how flexible computing infrastructure could accelerate data center interconnection and support power system reliability. Together, these results illustrate how AI infrastructure can evolve from static electricity consumption to dynamic participation in power system operation.

\section{Background and Prior Work}
\subsection{Data Center Demand Response}

Demand response programs have long explored the use of large electricity consumers as flexible resources for grid stability. 
Prior work has investigated several mechanisms for introducing flexibility into data center operations. In \textbf{temporal workload shifting}, delay-tolerant jobs are scheduled at different times to reduce peak electricity demand or respond to changing grid conditions \cite{liu2013}. Another approach leverages \textbf{geographically distributed data centers}, where workloads are migrated across locations to exploit regional differences in electricity prices, grid congestion, or renewable energy availability \cite{liu2011}. A related body of work investigates \textbf{carbon-aware computing}, in which workloads are scheduled according to the carbon intensity of the electricity grid \cite{radovanovic}. Finally, several studies have examined \textbf{data center demand response} in high-performance computing environments, where CPU-based batch workloads are delayed or throttled during grid stress events \cite{zhang}.

While these approaches demonstrated promising results in simulation or limited deployments, most focused on CPU-based systems or limited peak-demand curtailment scenarios rather than real-time GPU-based grid-interactive operation \cite{zhang}.

\vspace{-0.1in}
\subsection{AI-Specific Flexibility}

Recent studies have shown that AI clusters may provide significantly greater flexibility than traditional data center workloads. Training jobs often consist of long iterative processes with natural pause points such as checkpoint intervals, creating opportunities to temporarily slow down or suspend jobs without compromising correctness, since distributed training systems support checkpointing and preemption that allow jobs to resume from saved states \cite{shukla}.

Modern GPU clusters provide multiple mechanisms for controlling power consumption at both the hardware and workload levels. GPUs expose device-level power management interfaces that enable rapid reductions in instantaneous power draw, while cluster schedulers and AI workload managers can modulate power by prioritizing or pausing lower-priority jobs, resizing resource allocations, or adjusting training configurations dynamically. Recent work has shown that jointly tuning GPU power limits and training configuration parameters can substantially reduce the energy consumption of deep learning workloads while maintaining model accuracy and training performance \cite{you2023}.

Recent experimental demonstrations further showed that AI clusters can reduce power consumption in response to grid dispatch events while continuing to execute AI workloads and maintaining service-level agreements for high-priority tasks \cite{colangelo}.

\vspace{-0.1in}
\subsection{Key Differentiators in This Work}
While previous work demonstrated the feasibility of peak-demand curtailment, several important dimensions remain underexplored (see Table~\ref{tab:grid_services}).

\begin{table}[h]
\centering
\caption{Comparison of Data Center Flexibility Studies}
\label{tab:grid_services}
\renewcommand{\arraystretch}{1.5}
\begin{tabularx}{\linewidth}{@{} l >{\raggedright\arraybackslash}X >{\raggedright\arraybackslash}X @{}}
\toprule
\textbf{Dimension} & \textbf{Earlier Studies} & \textbf{This Article} \\
\midrule
Demonstration & Peak demand curtailment & Multiple grid services \\
Control scope & Single cluster & Multi-data-center systems \\
Flexibility mechanisms & Workload throttling & Throttling + geo-shifting \\
Grid signals & Scheduled events & Scheduled events + real-time dispatch + carbon signals \\
\bottomrule
\end{tabularx}
\end{table}


Previous data center demand response studies primarily focused on scheduled peak-demand events, simulation-based evaluation, or single-site workload throttling. In contrast, this work demonstrates software-defined flexibility in production-scale GPU clusters across multiple operational modes, including real-time grid dispatch, sustained curtailment, carbon-aware operation, and geographically distributed load shifting, while preserving service levels for priority jobs.

\section{Architecture for Power-Flexible AI Infrastructure}
Power-flexible AI infrastructure requires tight coordination between power systems and computing infrastructure. To enable this capability, data center operators must integrate grid signals, workload scheduling, and real-time power telemetry into a unified control framework that dynamically adjusts cluster power consumption without disrupting critical services.

\begin{figure*}[!t]
\centering
\includegraphics[width=\textwidth]{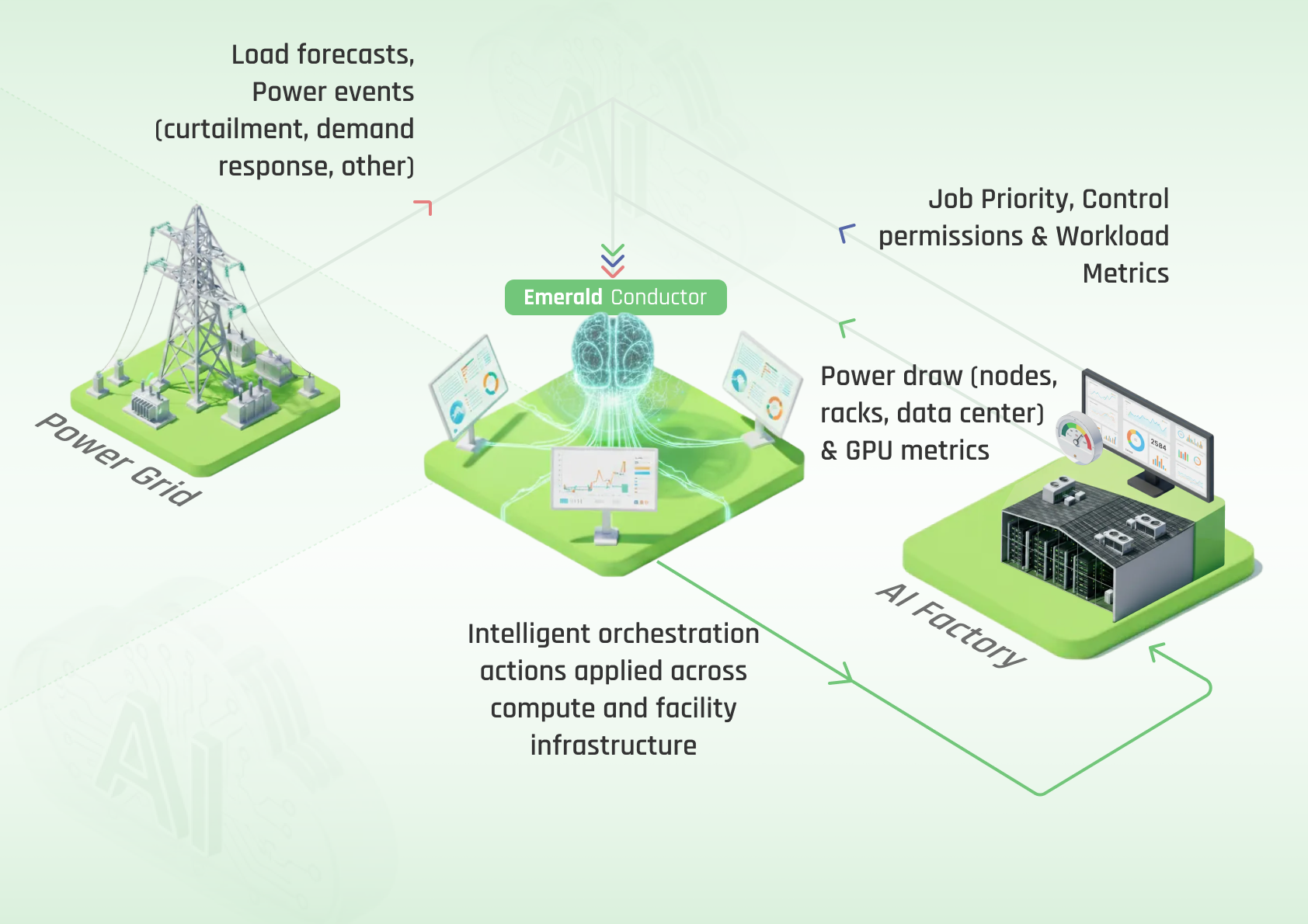}
\caption{Architecture of the grid-aware workload orchestration platform used in this study (Emerald Conductor).}
\label{conductor}
\end{figure*}

Figure \ref{conductor} illustrates the architecture we designed for our experimental deployment.
The architecture includes three primary components: a grid interface that receives dispatch signals from external power system operators, a workload orchestration layer that determines how computing jobs respond, and a telemetry-driven power modeling system that estimates and predicts cluster power consumption.

A software control layer connects the power system and the AI cluster. In our implementation, the Conductor platform performs this role by translating grid signals into cluster-level control actions such as GPU power capping and job scheduling adjustments.

\vspace{-0.1in}
\subsection{Grid Interface}
Our grid interface enables communication between the data center and the external power system actors.
Utilities or system operators send dispatch signals that describe real-time or anticipated grid conditions. 
The requested dispatch may represent demand response events, emergency load curtailments, congestion alerts, or carbon intensity indicators.
Each dispatch signal specifies power reduction target, event start time and duration, and ramp down/up requirements.


Dispatch events may arrive with advance notice or require immediate response. The grid interface converts these instructions into actionable power targets that the computing infrastructure must achieve within the specified response window.

\vspace{-0.1in}
\subsection{Workload Orchestration}



Our workload orchestration engine determines how the cluster should adjust its operation to meet requested power targets. The orchestration layer maintains a real-time view of cluster activity, including active jobs, resource allocations, and job priorities, and selects control actions that reduce power consumption while preserving application service levels. Typical control actions include GPU power capping and pausing or delaying lower-priority jobs.

AI clusters run heterogeneous workload mixes with different sensitivity to delays or slowdowns. Latency-sensitive inference services require stable performance, while many training jobs tolerate temporary slowdowns without affecting final model accuracy.

To support scheduling decisions, the orchestration engine in this deployment integrates directly with SLURM, leveraging existing job priority schemes to classify workloads into \textbf{flexibility tiers}. Critical jobs maintain strict performance guarantees, while background training or batch inference tasks provide the flexibility needed to meet grid power targets.

\vspace{-0.1in}
\subsection{Telemetry and Power Modeling}



Reliable power control relies on accurate visibility. Second-level GPU-level telemetry and independent rack-level power measurements feed our predictive models, which estimate cluster power consumption and anticipate the effects of workload adjustments. Over time, the controller builds a library of job power signatures that allows it to predict the impact of different control strategies. This enables rapid convergence on effective power allocation decisions while maintaining power and performance constraints.

\section{Experimental Demonstration}
We evaluated grid-responsive AI infrastructure using a GPU cluster operating under production conditions. Our experiment was conducted at a Nebius AI Factory in London, UK, using a \textbf{130 kW AI cluster with 96 NVIDIA Blackwell Ultra GPUs} connected through a high-speed interconnect network. The cluster ran continuously for five days while executing a diverse mix of AI workloads representative of modern training and inference pipelines.

During this period, grid partners (EPRI and National Grid) sent 22 dispatch events to the cluster. These events were derived from historical real-world examples, including peak load reduction based on ``TV Pickup'' demand patterns, emergency load reduction based on a 2019 lightning strike, and other incidents experienced by National Grid. Based on these events, test scenarios were developed using the EPRI DCFlex\footnote{https://dcflex.epri.com/} Flex MOSAIC™ framework\footnote{https://dcflex.epri.com/flex-mosaic}, which provides a structured approach for classifying large load flexibility. 

\vspace{-0.1in}
\subsection{Cluster Workloads}
The experimental cluster executed a mix of representative AI workloads designed to mimic production environments.
These workloads included:
\begin{itemize}
    \item large language model fine-tuning tasks
    \item multimodal training jobs
    \item batch inference pipelines
\end{itemize}

Examples included training runs based on models from the Llama family and other large-scale architectures. The cluster scheduler launched new jobs whenever GPU resources became available.
This scheduling strategy created a continuously evolving workload mix in which jobs frequently started and completed throughout the experiment, producing a power profile typical of real-world AI clusters.
The power control system had \textbf{no advance knowledge of the job schedule} (i.e., arrivals or completions), ensuring that the experiment reflected realistic operational conditions.

\vspace{-0.1in}
\subsection{Grid Dispatch Scenarios}
Grid partners submitted dispatch events via a remote portal where they specified power event parameters.
Some events included several minutes of advance notice, while others arrived with \textbf{zero warning}, forcing the cluster to respond immediately.

Throughout the test campaign, the system repeatedly adjusted cluster power consumption while jobs continued to start and finish dynamically.

\vspace{-0.1in}
\subsection{Measurement and Telemetry}

To evaluate system behavior, we monitored power consumption using both GPU and rack-level telemetry.
The \textbf{NVIDIA System Management Interface (smi)} was used to measure GPU power at second-level granularity.
Nebius provided \textbf{minute- and 20-second level rack power meter data} to help validate the NVIDIA smi readings and train cluster power models.

We designed a predictive power model that integrated these measurements to estimate total rack power while accounting for additional components such as CPUs, networking equipment, and storage subsystems. The controller in Conductor then used this model to anticipate how workload adjustments would affect cluster power consumption and to maintain precise compliance with dispatch targets.

\section{Results: Grid Services from AI Clusters}
The experiments show that AI clusters can deliver several forms of flexibility relevant to modern power systems.
Throughout the experimental campaign, the system met all requested power targets in the tested dispatch scenarios (i.e., 200+ total power targets and ramp constraints), demonstrating precise and reliable control of the power consumption of the cluster.

\vspace{-0.1in}
\subsection{Peak Demand Mitigation}

Electricity demand often exhibits predictable surges driven by human activity.  The UK's ``TV pickup" phenomenon is a clear example:  during major televised events, millions of households simultaneously switch on electric appliances at broadcast breaks.

\begin{figure*}[!t]
\centering
\includegraphics[width=\textwidth]{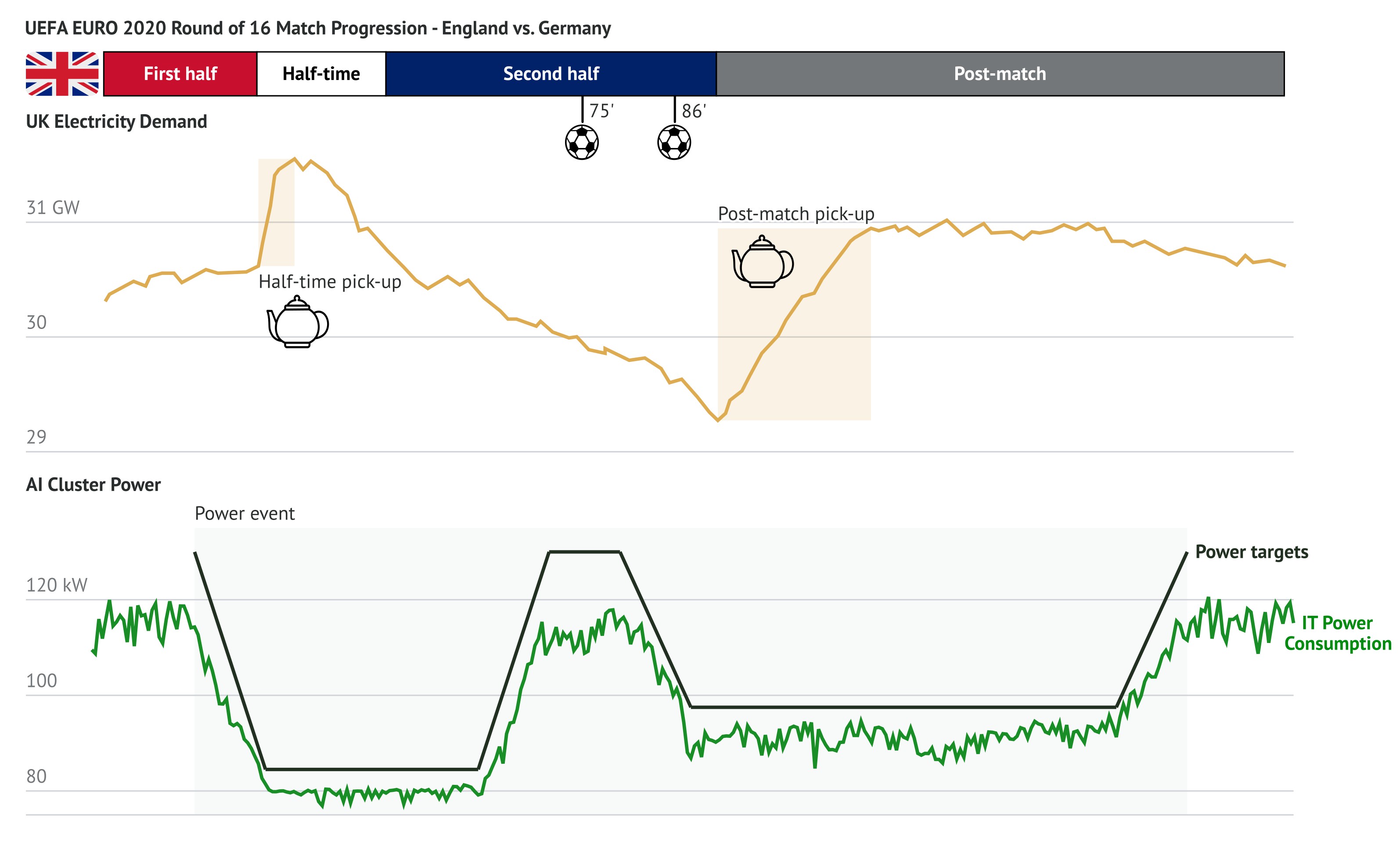}
\caption{AI cluster power response timed to offset a TV pickup (``tea kettle'') demand spike, overlaid with grid demand curves.}
\label{tv-pickup}
\end{figure*}

To emulate this scenario, we replayed a dispatch profile that replicated the demand spike observed during a major football game broadcast (Figure \ref{tv-pickup}).

As simulated residential demand increased, the AI cluster reduced its power consumption, creating an inverse power profile that compensated for the surge in demand. High-priority jobs maintained near-baseline throughput with negligible performance degradation.

\vspace{-0.1in}
\subsection{Emergency Load Reduction}
\begin{figure*}[!t]
\centering
\includegraphics[width=1.0\textwidth]{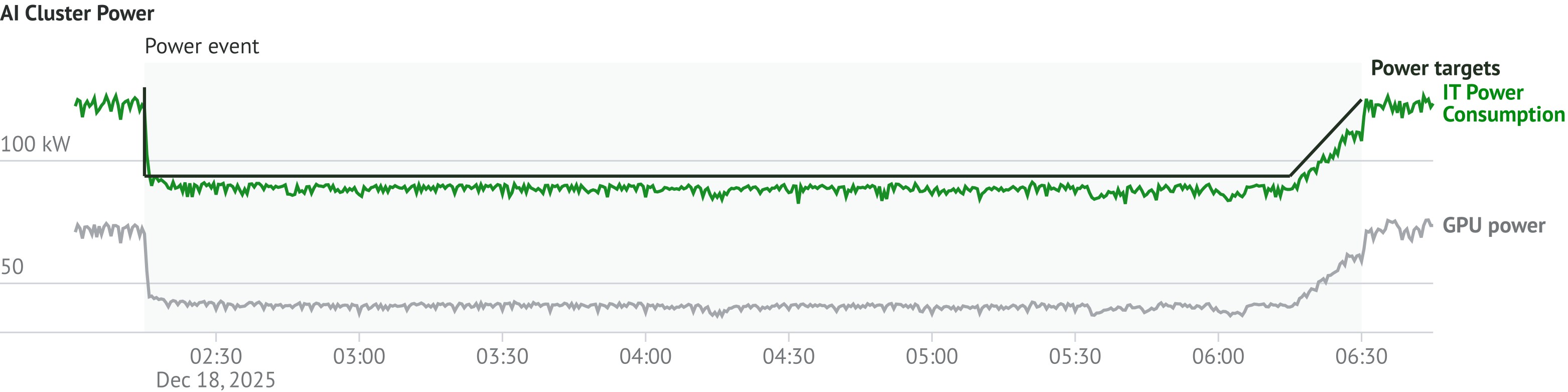}
\caption{AI cluster responds to a historical replay of a 2019 UK grid contingency associated with a major lightning event.}
\label{lightning-strike}
\end{figure*}

Power systems must also respond to unexpected generation losses or transmission faults. To test this capability, grid partners issued a surprise emergency dispatch modeled after a historical grid contingency that involved sudden generation loss (see Figure \ref{lightning-strike}).
The cluster reduced its power consumption by approximately \textbf{30\% within 40 seconds}, demonstrating response speeds comparable to some reserve, emergency demand response, and other fast grid support programs.

In additional experiments, the cluster achieved \textbf{40\% load reduction within approximately a minute} while maintaining performance guarantees for critical jobs.

\vspace{-0.1in}
\subsection{Sustained Curtailment}
Many grid stress events persist for extended periods rather than occurring as short disturbances.
To evaluate sustained flexibility, grid partners requested power reductions between 10\% and 40\% for durations ranging from two to ten hours. The controller delayed flexible jobs to off-peak periods while critical jobs continued operating.

\begin{figure*}[t]
\centering
\includegraphics[width=\textwidth]{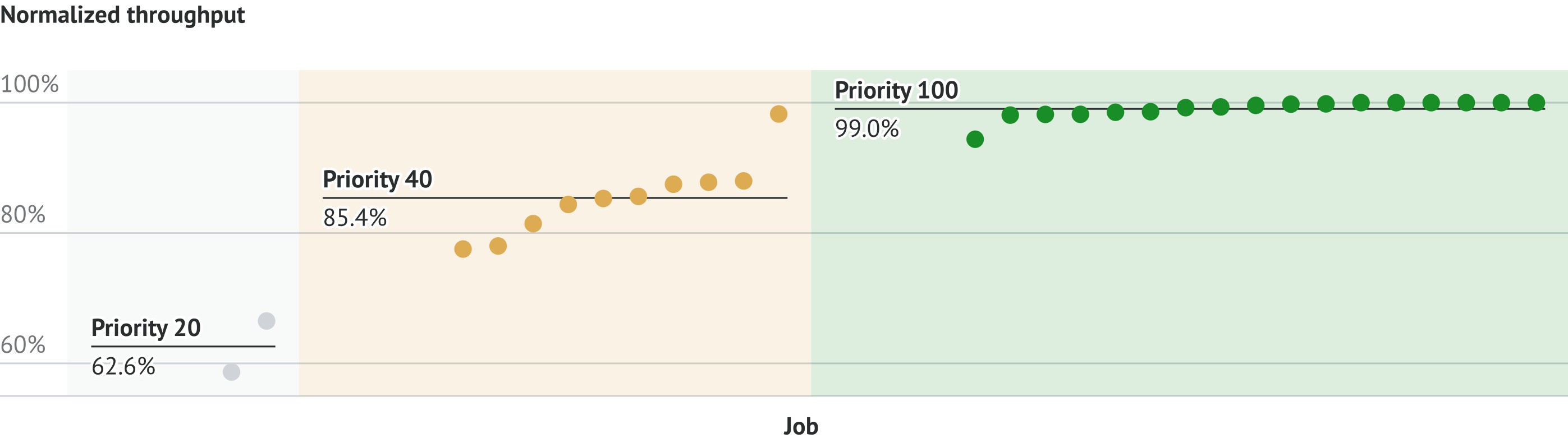}
\caption{Workload performance for 10-hour experiment, showing preservation of performance (measured as normalized throughput) for highest priority jobs.}
\label{job-performance}
\end{figure*}

As shown in Figure \ref{job-performance}, performance measurements reveal that high-priority jobs maintained nearly full throughput during the extended curtailment window. 
This is because our orchestration framework uses job priority metadata to preferentially target lower-priority jobs for power capping or delaying, while preserving performance for mission-critical jobs.

\vspace{-0.1in}
\subsection{Repeated Dispatch Events}
\begin{figure*}[!t]
\centering
\includegraphics[width=\textwidth]{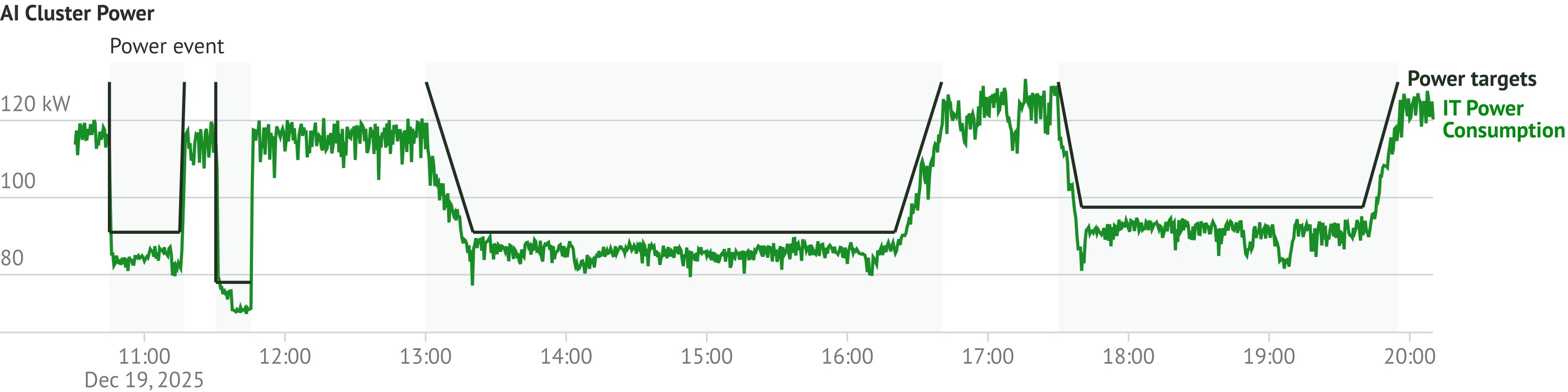}
\caption{AI cluster power response to live grid events submitted by National Grid and EPRI, including demonstrating 	$<$40-second and 1 minute response times to zero-notice, immediate-ramp-down events.}
\label{epri-experiment}
\end{figure*}

Grid operators frequently issue multiple dispatch instructions within short intervals. To replicate these conditions, grid partners submitted several dispatch signals within a ten-hour window without prior coordination, as shown in Figure \ref{epri-experiment}.

The system executed all requested power adjustments, including rapid ramp-down events with response times on the order of tens of seconds.
These results indicate that AI clusters can operate as \textbf{continuously dispatchable grid resources} capable of responding repeatedly to evolving grid conditions.

\vspace{-0.1in}
\subsection{Carbon-Aware Operation}
\begin{figure*}[t]
\centering
\includegraphics[width=\textwidth]{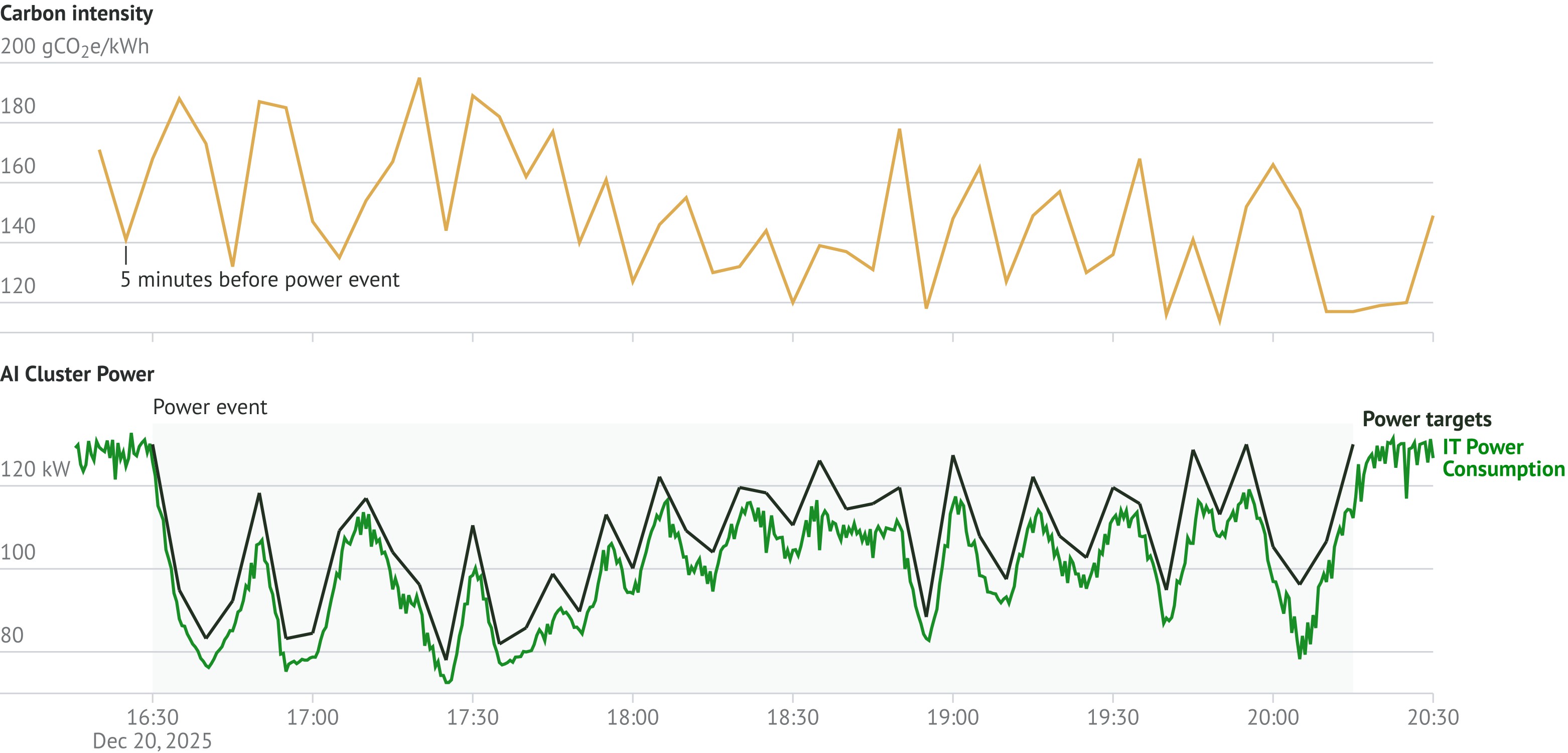}
\caption{Carbon-aware load following and power tracking across grid conditions.}
\label{carbon-aware}
\end{figure*}

Electricity system carbon intensity fluctuates throughout the day as renewable output and generation dispatch change. In one experiment (shown in Figure \ref{carbon-aware}), the cluster followed a \textbf{five-minute carbon intensity signal} derived from real grid conditions.

The controller reduced power consumption during higher-emission periods and increased utilization when cleaner electricity became available. This capability allows computing workloads to align electricity demand with renewable availability and reduce the carbon footprint of AI infrastructure.

\section{Geo-Load Shifting Across Data Centers}
While local power modulation provides a critical dimension of flexibility, large-scale AI deployments often span geographically distributed data centers.
This distribution introduces an additional mechanism for shifting
computational load across regions in response to localized grid constraints. The following results demonstrate geographical load shifting between two active data centers in response to a grid power event.  Testing occurred at a data center within Dominion Energy’s service territory and was enabled by leveraging EPRI’s DC Flex collaborative and its members—NVIDIA, PJM, Constellation Energy, ComEd, and Exelon.

\vspace{-0.1in}
\subsection{Load Shifting Architecture}




We deployed an inference-serving architecture that routed load across two geographically distributed clusters using real-time performance metrics. The multi-region setup relied on multi-region load balancing, latency-aware routing logic, and shared model hosting.

Regional inference clusters and a central load balancer formed the core architecture. The inference clusters ran NVIDIA Dynamo on Kubernetes, utilizing vLLM as the inference engine. Each vLLM worker used a single GPU serving the Qwen2.5-32B-Instruct model, and both clusters operated in aggregated serving mode. An Envoy load balancer routed requests between clusters 
based on total request latency.

To simulate realistic user demand, we leveraged NVIDIA’s AIPerf benchmarking tool to generate a continuous and measurable inference load across the system.

Power telemetry was collected at the GPU and rack level to ensure that the applied power limits were strictly respected. The specific telemetry approach involved:
\begin{itemize}
    \item Collecting GPU power measurements at five-second granularity using NVIDIA’s DCGM-Exporter;
    \item Validating readings against minute-level rack power data provided by Oracle.
\end{itemize}

\subsection{An Inference Load Shift Demonstration}
This demonstration evaluated the impact of restricting the power utilization of one inference cluster while allowing load to fail over to another cluster. We used two identically configured Oracle data centers in Ashburn, VA, and Chicago, IL, each hosting a 60~kW cluster with 80 NVIDIA H100 GPUs.

While maintaining a constant baseline load on both clusters, we curtailed power consumption in the Ashburn cluster by enforcing GPU power limits and removing GPUs from the active inference pool. We then monitored the resulting load shift to Chicago and evaluated user-facing metrics including Time to First Token (TTFT) and end-to-end request latency.

\begin{figure*}[!t]
\centering
\includegraphics[width=0.95\textwidth]{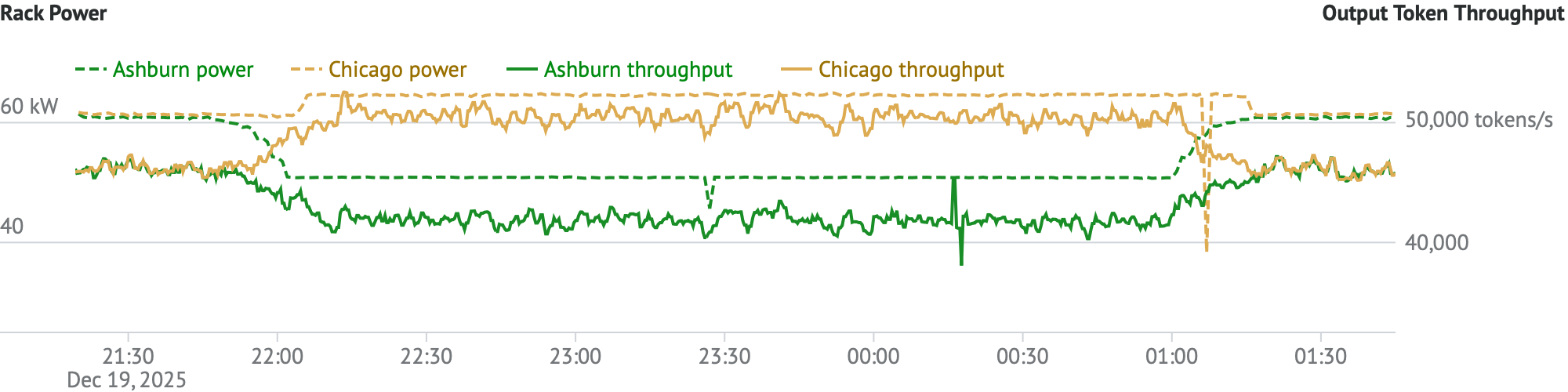}
\caption{Rack power consumption and tokens/second throughput in the Chicago and Ashburn inference clusters. The throughput for the clusters diverges during the power reduction event in the Ashburn cluster.}
\label{geo-combined}
\end{figure*}

During a targeted experiment, we set a 375 W power limit on all GPUs within the Ashburn cluster. As shown in Figure \ref{geo-combined}, the cluster executed a 15-minute ramp-down from full power, followed by a sustained three-hour hold at the new limit. Concurrently, the Envoy load balancer detected the constrained capacity in Ashburn and automatically rerouted traffic. This resulted in a 3.1 kW increase in power consumption at the Chicago cluster as it absorbed the displaced load.

The rerouting of traffic had a direct and measurable effect on cluster throughput. As illustrated in Figure \ref{geo-combined}, while the Ashburn cluster remained power-constrained, its token generation rate proportionally decreased. Simultaneously, the Chicago cluster experienced an equivalent surge in tokens-per-second throughput,
consistent with successful transfer of a portion of inference load.

We analyzed Time to First Token (TTFT) during the power event to quantify the user-facing impact of this power constraint and subsequent load shift.
In the power-restricted Ashburn cluster, the imposed power constraint caused a sustained but manageable increase of approximately 30 ms in average TTFT. Conversely, the Chicago cluster experienced only a brief, temporary spike in TTFT as the influx of new requests arrived. This spike rapidly subsided once the cluster's autoscaler provisioned additional GPU capacity to handle the newly shifted load from Ashburn.


\section{Discussion}
The experimental results demonstrate that AI clusters can modulate power consumption across multiple timescales while continuing to execute production workloads. Given these capabilities, 
power-flexible AI data centers may influence how grids connect new loads, maintain reliability, and integrate renewable energy resources.

\vspace{-0.1in}
\subsection{Faster Data Center Interconnection}
Power-flexible computing infrastructure has important implications for how grid operators evaluate new large electricity loads. Today, planners typically treat data centers as firm loads, meaning the grid must guarantee full power delivery at all times, including during rare grid system stress events. This assumption stems from a lack of data on the flexibility capabilities of large computing loads, and forces utilities to design transmission and generation infrastructure for worst-case demand occurring at the same time across loads.

If data centers can reliably reduce power consumption during stressed grid operating conditions, operators may instead allow them to connect under \textbf{alternative or non-firm agreements} that permit temporary curtailment during periods of system stress.\footnote{A December 2025 Federal Energy Regulatory Commission order resulted in some ISOs revising their tariffs regarding Non-Firm Contract Demand Transmission Service.
This may allow some data centers to apply as non-firm loads.} Because such arrangements reduce or defer the need for infrastructure upgrades identified for extreme conditions and system contingencies, they can shorten interconnection timelines and result in lower system costs. Power-flexible computing therefore provides a pathway to accelerate the deployment of new AI infrastructure while preserving grid reliability.

\vspace{-0.1in}
\subsection{Grid Reliability}
Beyond enabling faster interconnection, power-flexible AI clusters can also support grid reliability by operating as controllable demand resources. In our experiments, the system responded to dispatch signals within tens of seconds, sustained reduced power consumption for extended periods, and repeatedly followed dispatch instructions under dynamic conditions.

Traditionally, grid operators rely primarily on generation resources or energy storage to balance supply and demand. Flexible computing loads introduce an additional option: controllable demand that can reduce consumption during periods of system stress or unexpected disturbances. Because many AI workloads can tolerate controlled changes in execution speed, these services can often be provided without abrupt shutdowns, distinguishing AI clusters from many traditional industrial demand response resources.



\vspace{-0.1in}
\subsection{Renewable Integration}

These tests also indicate that power-flexible computing can support renewable energy integration. Wind and solar generation fluctuate over time, sometimes creating periods in which supply exceeds local demand and renewable generation must be curtailed, often due to transmission congestion 
\cite{perry}.
Many AI workloads, particularly large training jobs, can be scheduled flexibly across time.
By shifting flexible AI jobs to periods with abundant renewable energy, data centers can absorb electricity that might otherwise be curtailed, improving clean energy utilization and potentially lowering the carbon footprint of AI infrastructure. If data centers can also reduce consumption during periods of limited renewable generation, they may reduce the need for some transmission or firm capacity upgrades while improving system flexibility.

\vspace{-0.1in}
\subsection{Implications for Data Center Design}
Power-flexible computing may influence future data center design by encouraging tighter integration of power-aware scheduling into cluster management systems and software interfaces that respond to grid signals in real time. At larger scales, orchestration platforms may coordinate workloads across multiple facilities, combining local power modulation with geo-shifting to respond to grid conditions, carbon intensity, and renewable availability.

Scaling power-flexible AI infrastructure to large deployments introduces challenges, as modern AI factories include thousands of GPUs along with complex electrical and thermal systems. Effective power control therefore requires coordinated management across computing, cooling, and facility power infrastructure.

\section{Conclusion}
AI infrastructure growth is placing increasing pressure on power systems, where data centers are traditionally modeled as inflexible loads, causing costly upgrades and interconnection delays.
This article demonstrates how modern GPU-based AI data centers can operate as flexible, grid-interactive resources. Through software-based workload orchestration, clusters can dynamically adjust power consumption across multiple timescales while preserving application service levels for priority workloads. Beyond prior demonstrations of peak-demand curtailment, the capabilities described here include rapid response to grid events, sustained load reduction, carbon-aware operation, and geographic workload shifting across multiple data centers. These mechanisms transform AI infrastructure from a passive electricity consumer into an active participant in power system operation.


\section*{Acknowledgments}
This paper benefited greatly from input and review by collaborators, including:
\begin{itemize}
    \item T. Bedard, D. Porter, and A. Ratnayake at \textbf{EPRI};
    \item S. Smith, P. Tagare, S. Brown, D. Adkins, N. Moturi, J. Avramidis, A. Badcock-Broe, M. Cochrane, G. Kelly, N. Kesharaju, L. Ogilvie, C. Perry, A. Roache, P. Thorp, and R. Woodward at \textbf{National Grid};
    \item M. Boroditsky, M. Horvath, P. Morley, D. Mukhortova, A. Miller, I. Znamenskii, and M. Welsh at \textbf{Nebius};
    \item N. Caprez, J. Parker, V. Troy, M. Spieler, A. Yousufzai, Z. MacFarland, and K. Sarkisian at \textbf{NVIDIA};
    \item J. Jackson, R. Ramaswamy, S. Chen, and R. Caputo at \textbf{Oracle}; and
    \item M. Beams, S. Jain, J. Megrue, and A. Vijaykar at \textbf{Emerald AI}.
\end{itemize}

\def\refname{REFERENCES}

\end{document}